\documentclass[aps,preprint,prb,showpacs,groupedaddress]{revtex4}
\usepackage{graphicx}

\begin{document}


\title{Ab-initio investigation of phonon dispersion and anomalies in palladium}


\author{Derek A. Stewart}
\email[]{stewart@cnf.cornell.edu}
\affiliation{Cornell Nanoscale Science and Technology Facility, \\ Cornell University, Ithaca, New York, 14853, USA}


\date{\today}

\begin{abstract}
In recent years, palladium has proven to be a crucial component for devices ranging from nanotube field effect transistors to advanced hydrogen storage devices.  In this work, I examine the phonon dispersion of fcc Pd using first principle calculations based on density functional perturbation theory.  While several groups in the past have studied the acoustic properties of palladium, this is the first study to reproduce the phonon dispersion and associated anomaly with high accuracy and no adjustable parameters.
\end{abstract}

\pacs{63.20.D,71.18,71.20}

\maketitle

\section{Introduction}
Recent interest in hydrogen storage systems and nanoscale devices has highlighted the crucial role palladium plays in a wide range of systems.  Hydrogen sensors based on Pd nanowires\cite{hydrogen_sensor_2001} show both fast response and low power requirements.  Recent experimental work indicates that Pd leads provide ohmic contacts for nanotube field effect transistors, a feature crucial for large scale device integration\cite{javey}.  In addition to nanoscale applications, bulk palladium also presents interesting properties that have fascinated researchers for years.  It possesses a high magnetic susceptibility and sits on the edge of magnetism.  Perhaps due to this, no measurable superconductivity has been found in the system for any temperature.

Phonon scattering events play a crucial role in both superconductivity in bulk systems and transport in nanoscale interconnects.  In this work, I examine the phonon dispersion of palladium from a first principles perspective.  Recent advances in density functional perturbation theory (DFPT) have made it possible to examine the acoustic properties of materials at a level of accuracy previously reserved only for electronic properties\cite{baroni1987,dfpt_review}.  Prior to the development of this approach, researchers were forced to use  a frozen phonon technique that required large supercells or phenomenological approaches that relied on numerous fitting parameters.

Interest in the acoustic properties of palladium grew after Miiller and Brockhouse examined the phonon dispersion of palladium in detail using inelastic neutron scattering\cite{brockhouse1968,brockhouse1971,miiller1975}.  They observed a change in the slope of the [$\zeta\zeta$0] transverse acoustic, TA$_{1}$, branch of the dispersion curve around $q$-vector $\frac{2\pi}{a}$[0.35, 0.35, 0] that differed from phonon anomalies observed in other metals.  The fcc lattice constant in this case is given by $a$.  The anomaly extends over a broad range of wave vector space and also decreases rapidly with temperature.  These features initially cast doubt on whether this feature was a Kohn anomaly brought about by nesting within parallel sections of the Fermi surface\cite{kohn}.  

For small q-vectors in metals, itinerant electrons are able to effectively screen the positive ionic charge revealed by lattice vibrations.  However once the phonon wavevector, $\bf{q}$, spans different Fermi sheets and links different electronic states, $\bf{k_2}=\bf{k_1}+\bf{q}$, the electrons are no longer able to effectively shield the induced positive charge and the acoustic properties undergo a marked change, the Kohn anomaly\cite{kohn}.  Making use of Fermi surface data from augmented plane wave calculations, Miiller found transition vectors between parallel sheets of the fifth band hole surface that supported the presence of a Kohn anomaly\cite{miiller1975} in Pd.  Later calculations based on the generalized susceptibility, $\chi_{0}\left(q\right)$ in Pd also indicated that nesting between sheets of the fifth band could contribute to the extended anomaly\cite{freeman1979}.  This work also predicted the presence of a weaker Kohn anomaly at $\bf{q}=\frac{2\pi}{a}$[0.28, 0.28, 0.28] in the TA$_{1}$ branch along the [111] direction.  However, Maliszewski \textit{et. al.} noted that while the sharp peak in $\chi_{0}\left(q\right)$ does occur in the same wave vector range as the [110] phonon anomaly, it does not correspond to the behavior of the experimental anomaly in the [110] direction\cite{maliszewski1979}.  They suggested that spin fluctuations or paramagnons could be playing a crucial role in the structure of the [111] anomaly.    

Several phenomenological approaches have provided fitting routines to generate the palladium phonon dispersion.  These include studies using a six parameter screened shell model\cite{fielek1980}, a three parameter Morse potential approach\cite{merz1984}, and a four parameter Coulomb potential technique\cite{Gupta1985} among others.  While these approaches can provide general agreement with the phonon dispersion of palladium, it should be noted that these techniques do not reproduce the phonon anomaly in the [110] direction.  The large number of fitting parameters inherit in these schemes also limits the predictive ability of these approaches.  One study based on a pseudopotential approach with a short range pair potential for $d$-$d$ orbital interactions did show evidence for a phonon anomaly in the TA$_{1}$ branch in the [110] direction\cite{antonov1991}.  However, the predicted anomaly was shifted to smaller phonon wavevectors and showed a much more dramatic phonon softening than the measured anomaly. 

Two recent works have examined acoustic properties of palladium in the framework of density functional perturbation theory.  Savrasov and Savrasov looked at electron-phonon interactions for a range of materials using a full potential linear muffin tin orbital approach (FP-LMTO)\cite{savrasov1996}.  Their calculations reproduce the general features of the experimental phonon dispersion of palladium, however, the phonon wavevector sampling used was too coarse to reveal phonon anomalies.  A recent estimate of the superconducting transition temperature for palladium under pressure also made use of phonon dispersion curves and did not observe any phonon anomalies\cite{takezawa}.  

In this work, I present high resolution calculations for the phonon dispersion of fcc palladium.  I compare direct calculations at q-vectors along high symmetry lines with results based on interatomic force constants.  In particular, I focus on the phonon anomaly in the [110] direction and show that DFPT is able to reproduce this feature with high fidelity.  Since this anomaly is due to phonon-electron interactions, I also determine the location of the Kohn anomaly based on Fermi surface analysis and the two different techniques show good agreement.  I have also examined the phonon dispersion along the [111] direction and I do not find evidence for the phonon anomaly previously predicted by Freeman \textit{et. al.}\cite{freeman1979}.

\section{Density Functional Perturbation Theory}
For a periodic lattice of atoms, the only portions of the crystal Hamiltonian that depend on the ion positions are the ion-electron interaction terms, $V_{R}(\textbf{r})$, and the ion-ion interaction term, $E_{ion}(\textbf{R})$.  The force on a given atom consists of a component that is responding to the surrounding electron charge density, $n(\textbf{r})$, and another term that accounts for ion-ion interactions.  The interatomic force constants used to determine phonon frequencies in the system are found by differentiating the force, F$_{I}$ on a given ion, $I$, by the change in position of ion $J$.

\begin{eqnarray}
C_{R_{I},R_{J}} 
=\frac{-\partial F_{I}}{\partial R_{J}}
=\int \frac{\partial n(\textbf{r})}{\partial R_{J}}\frac{\partial V_{R}(\textbf{r})}{\partial R_{I}}d\textbf{r} \nonumber \\
+\int n(\textbf{r})\frac{\partial^2 V_{R}(\textbf{r})}{\partial R_{I}\partial R_{J}}d\textbf{r} +\frac{\partial^2 E_{ion}(\textbf{R})}{\partial R_{I}\partial R_{J}}
\end{eqnarray}

The interatomic force constant matrix (IFC) above consists of seperate electronic and ionic contributions.  The ionic contribution can be related to an Ewald sum and calculated directly\cite{giannozzi}.

There are two factors that are critical for the determination of the electronic contributions to the IFC.  The first is the ground state electron charge density, $n_{R}(\textbf{r})$.  The second is the linear response of the ground state electron density to a change in the ion geometry.  These quantities can be calculated directly within the density functional framework without resorting to fitting data from experiment.

While first principle calculations readily determine the equilibrium electron charge density, $n_{R}(\textbf{r})$, some additional effort is required to determine $\partial n_{R}(\textbf{r})/ \partial R_{I}$ and the corresponding IFC.  However, the computational effort required is on the same order as the standard ground state energy calculation.  Interatomic force constants for periodic structures can be determined through the use of density functional perturbation theory\cite{baroni1987,dfpt_review}.  In this approach, the Kohn-Sham equations for the charge density, self-consistent potential and orbitals are linearized with respect to changes in wave function, density, and potential variations.  For phonon calculations, the perturbation, $\Delta V_{ion}$, is periodic with a wave vector \textbf{q}.  This perturbation results in a corresponding change in the electron charge density, $\Delta n(\textbf{r})$.  Since the perturbation generated by the phonon is periodic with respect to the crystal lattice, we can Fourier transform the self-consistent equations for the first order corrections and consider the Fourier components, $\Delta n(\textbf{q}+\textbf{G})$, of the electron density change, where $\textbf{G}$ is any reciprocal lattice vector.  In k-space, this problem can now be completely addressed on a sufficiently dense k-point grid in the unit cell Brillouin zone and fast Fourier transforms can be used to convert back to the real space form.  This removes the need to examine frozen phonon supercells and greatly accelerates calculations.  This approach has been successfully applied to reproduce the phonon dispersion of materials ranging from semiconductors, metals, and multilayers (see the review by Baroni \textit{et. al}\cite{dfpt_review} for examples).
  
\section{Results}
\subsection{Choice of Pd Pseudopotential}
Total energy and phonon calculations were performed on fcc palladium using the first principles package Quantum-ESPRESSO\cite{pwscf}.  Several varieties of local density approximation (LDA) and generalized gradient approximation (GGA) pseudopotentials were considered in this study including ultrasoft Perdew-Burke-Ernzerhof (PBE) pseudopotentials\cite{pbe_gga} as well as Perdew-Zunger (LDA) ultrasoft pseudopotentials\cite{pz_lda}.  In both sets, the effect of including semi-core $d$ states was also considered. I found that all GGA pseudopotentials predicted a magnetic ground state for fcc Pd.  In addition, LDA pseudopotentials that did not take into account the semi-core $d$ state predicted a magnetic ground state as well.  Bulk palladium is not ferromagnetic, although previous works have indicated a small expansion of the crystal can induced a finite magnetic moment\cite{fcc_pd_mag_trans}.  The LDA pseudopotential that includes the semi-core $d$ state provided the physical paramagnetic palladium ground state and is used throughout the remainder of the study.  The inability for GGA Pd pseudopotentials to provide the correct non-magnetic ground state has also been recently found by other groups using different first principle approaches\cite{Alexandre2006,Sob2005}.  It is also interesting to note that a recent work examining the phonon dispersion in fcc palladium found good agreement with LDA pseudopotentials, but poor agreement with GGA pseudopotentials\cite{takezawa}.  Since the Pd GGA pseudopotential overestimates the equilibrium unit cell volume, this leads to an overall shift in the predicted phonon dispersion to lower frequencies.

\subsection{Lattice Dynamics in Palladium}
Total energy calculations made use of a 34 Ry plane wave cut-off and a 348 Ry energy cut-off for the electron density.  A 16x16x16 shifted Monkhorst-Pack k-point grid\cite{mp_grid} was used in all calculations and electron occupation was treated with a Methfessel-Paxton smearing parameter\cite{methfessel_paxton1989}, $\sigma$, of 0.01 Ry.  The smearing parameter plays the role of a fictitious electronic temperature and acts to smear out discontinuities found in the density of states at zero temperature.  The exact result is obtained in the limit as $\sigma \rightarrow 0$.  In most cases, the phonon frequencies are fairly insensitive to $\sigma$, however near Kohn anomalies I found that very small $\sigma$ values are often required.  

\begin{figure}
\begin{center}
\centering
\includegraphics[angle=0,width=12.50cm]{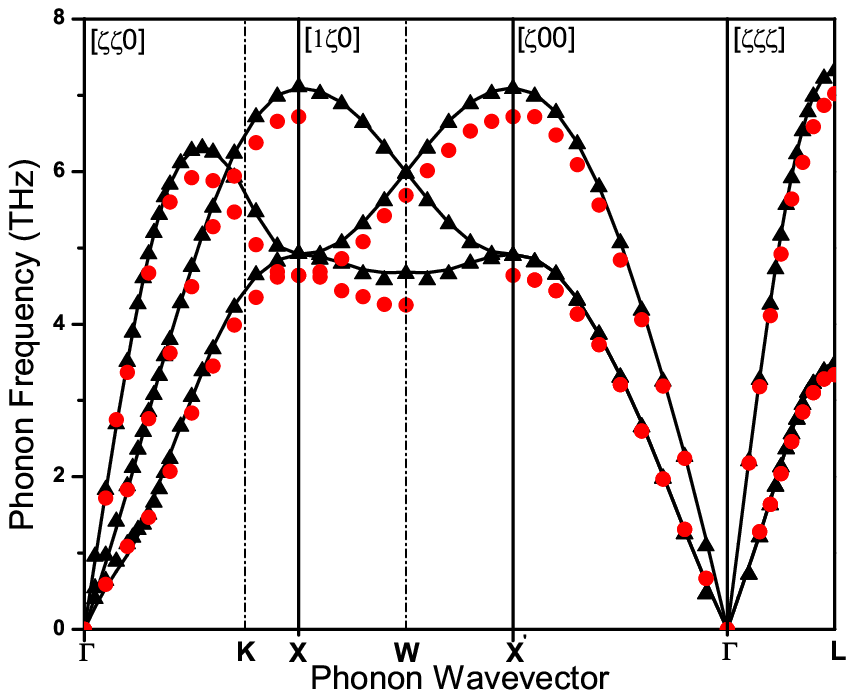}
\caption{}
\label{full_phonon_dispersion} 
\end{center} 
\end{figure}

Phonon frequencies were calculated directly for various phonon wavevectors along high symmetry lines using density functional perturbation theory and are shown in Figure ~\ref{full_phonon_dispersion}.  The phonon dispersion was also calculated based on the interatomic force constants obtained from a 8x8x8 Monkhorst-Pack $q$-vector grid.  The acoustic sum rule (ASR) was used for the interatomic force constant calculations, but this had a fairly small effect of the predicted phonon frequencies  For example, the maximum difference in phonon frequencies between calculations with and without ASR occurred at the $\Gamma$ point ($\Delta \omega = 0.07$ THz).  In the region of the [110] Kohn anomaly, the frequency difference was $0.003$ THz or less.  The direct and ASR interatomic force constant results are shown in comparison to phonon frequencies determined in a previous experimental study at 120 K\cite{brockhouse1971}.  The overall agreement for the direct phonon calculations is good with an average 5\% difference between predicted and measured values.  The 8x8x8 $q$-vector IFC results shows good agreement with the direct calculations for all of the symmetry lines, but have some difficulty reproducing experimental results at low frequencies in the acoustic branches.  I found coarser $q$-vector grids were unable to reproduce the experimental phonon dispersion curves.

It should be stressed that the phonon dispersion curves from both the direct calculations and interatomic force constants are calculated without any adjustable parameters and make no reference to experimental data.  There are two clear sources for differences between the measured and calculated phonon frequencies.  The first is due to the fact that experimental studies were performed at 120 K, while first principle calculations assume zero temperature.  Thermal expansion of the unit cell will act to shift the phonon frequencies lower.  In addition, the local density functional approximation (LDA) has a tendency to overbind and predict slightly smaller lattice constants than those observed experimentally.  Using LDA, the equilibrium lattice constant for fcc Pd was determined to be 3.88 \AA, while the measured lattice constant is slightly larger, 3.89 \AA.  The smaller predicted lattice constant will lead to a more compressed unit cell and a general shift to higher phonon frequencies.  This trend of overestimating the phonon frequency is greatest in the longitudinal acoustic branch.  

\begin{figure}
\begin{center}
\centering
\includegraphics[angle=0,width=12.50cm]{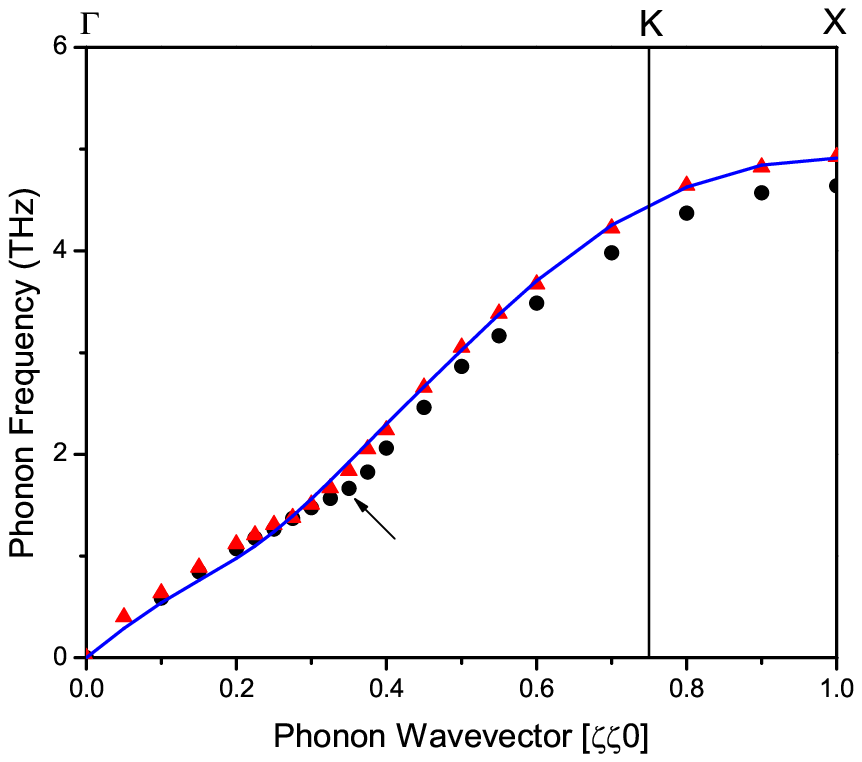}
\caption{}
\label{110_line} 
\end{center} 
\end{figure}

While there are some slight differences between the current work and previous experimental studies, it is important to note that the direct and IFC calculations replicate the phonon anomaly observed in the [110] direction.  Previous parameter based approaches have been unable to capture this anomaly.  This feature in the TA$_1$ branch is shown in Figure ~\ref{110_line} in comparison with experimental data at 8 K\cite{miiller1975}.  The slope of the calculated phonon dispersion changes at approximately $\bf{q}=\frac{2\pi}{a}$[0.30, 0.30, 0], indicating a Kohn anomaly.  The slope of the experimental phonon dispersion changes at a slightly higher q-vector, $\bf{q}=\frac{2\pi}{a}$[0.35, 0.35, 0].  This small difference in the calculated and experimental q-vector for the Kohn anomaly could be related to the smaller lattice constant predicted by density functional techniques.

\begin{figure}
\begin{center}
\centering
\includegraphics[angle=0,width=12.50cm]{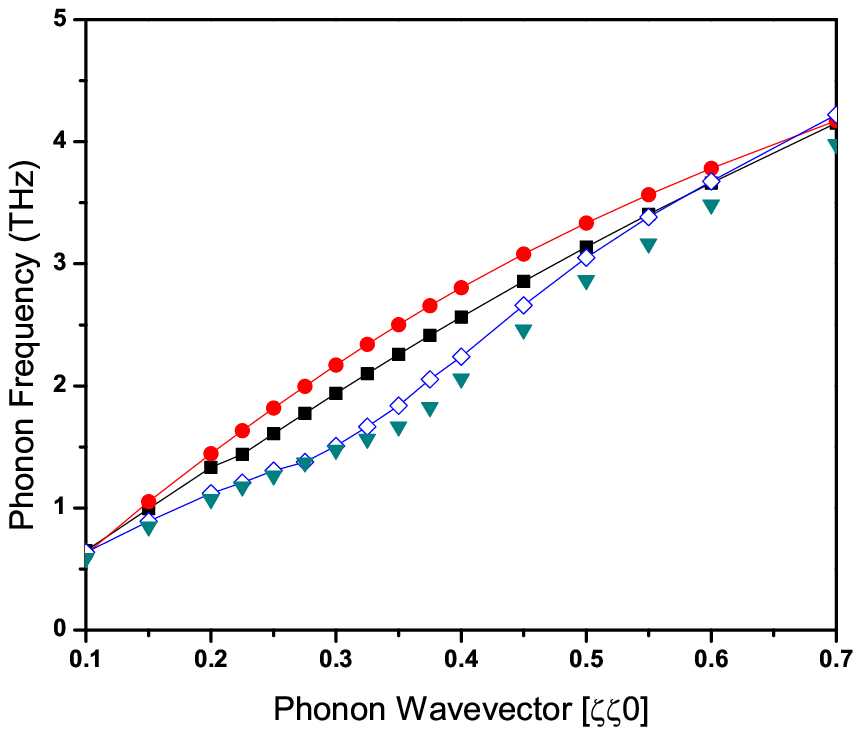}
\caption{}
\label{ta1_degauss} 
\end{center} 
\end{figure}

As an additional check to determine whether this feature is a Kohn anomaly, I examined the TA$_{1}$ branch as a function of the smearing parameter, $\sigma$.  Since this parameter serves as a fictitious temperature, increasing it will affect virtual transitions between occupied and empty states in an energy range $\sigma$ around the Fermi energy\cite{piscanec}.  This will govern how effectively electrons are able to shield motion in the ion lattice and should lead to the disappearance of the anomaly.  The TA$_{1}$ branch for smearing parameters ranging from 0.01 Ry to 0.10 Ry are shown in Figure ~\ref{ta1_degauss} in comparison with the experimental values.  It is clear that as the smearing parameter increases, the Kohn anomaly disappears.  For phonon wavevectors far from the Kohn anomaly, like $\bf{q}=\frac{2\pi}{a}$[0.70, 0.70, 0], adjusting the smearing parameter has little effect on the TA$_{1}$ phonon frequency.  

\begin{figure}
\begin{center}
\centering
\includegraphics[angle=0,width=7.5cm]{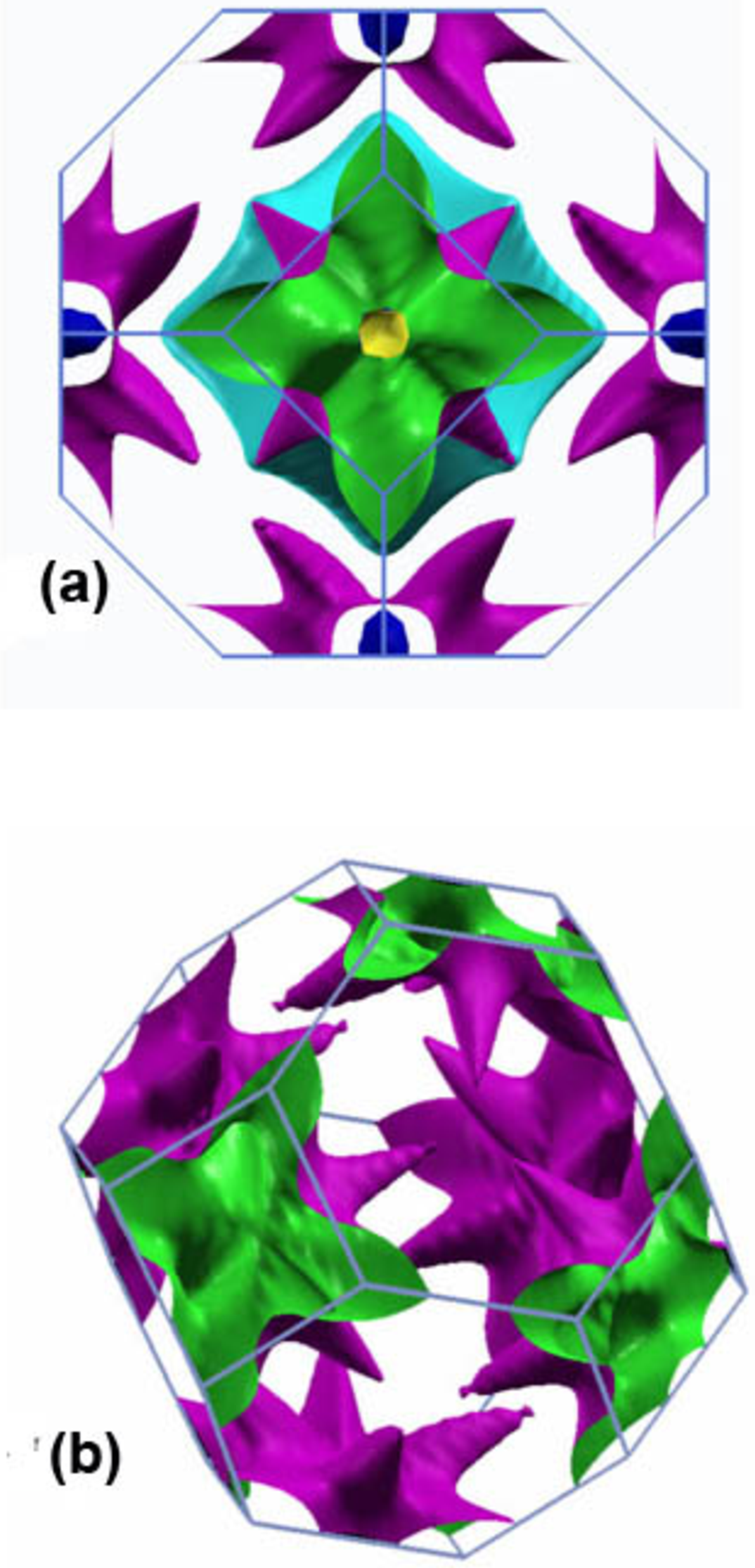}
\caption{}
\label{pd_fcc_fermi} 
\end{center} 
\end{figure}

The Fermi surface for fcc Pd was extracted from the calculation as well.  There are three bands that contribute at the Fermi energy (Fig. ~\ref{pd_fcc_fermi}(a)).  The fourth band barely overlaps with the Fermi energy and only exhibits small cusps at the center of the [100] faces.  The sixth band is centered in the Brillouin zone and although faceted, has high $s$ character and is similar to a free electron Fermi surface.  The Fermi surface for the $d$-like 5th band in fcc palladium is shown in Figure ~\ref{pd_fcc_fermi}(b).  

\begin{figure}
\begin{center}
\centering
\includegraphics[angle=0,width=12.50cm]{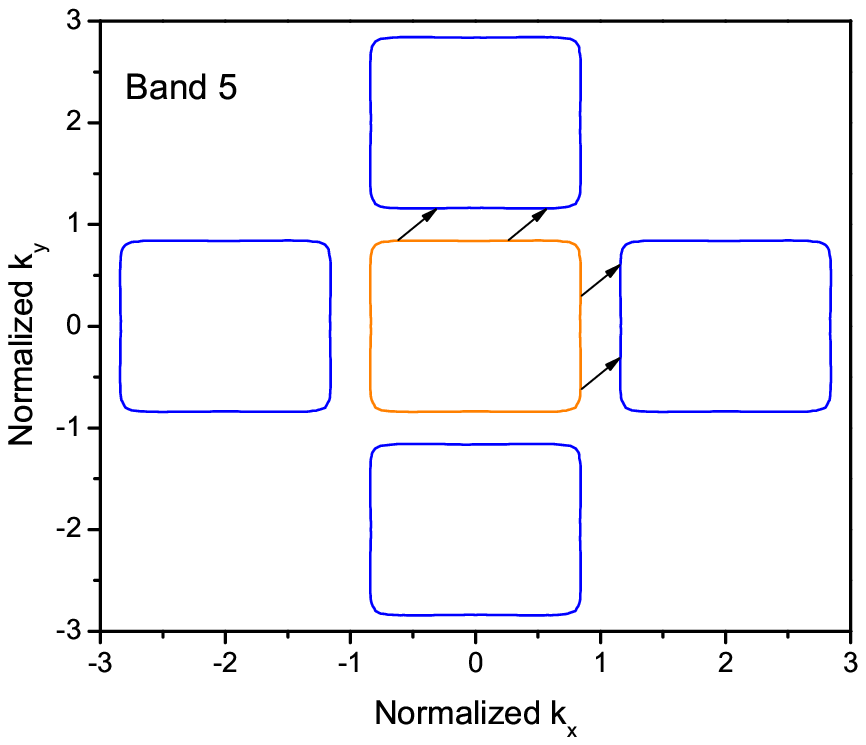}
\caption{}
\label{band5_crossx} 
\end{center} 
\end{figure}

The fifth band Fermi surface is dominated by wide sheets that sit in the [100] faces of the Brillouin zones.  These parallel sheets provide important contributions to the Kohn anomaly in the [110] direction.  This can be seen by examining the intersection of the fifth band with the $k_z = 0$ plane in the central Brillouin zone and neighboring Brillouin zones (Fig. ~\ref{band5_crossx}).  The fifth band Fermi surface forms squares in the $k_z = 0$ plane that can be linked with $\bf{q}=\frac{2\pi}{a}$[0.30, 0.30, 0] nesting vectors.  These nesting vectors occur at the phonon wavevector calculated for the Kohn anomaly in the [110] direction.   
  
\begin{figure}
\begin{center}
\centering
\includegraphics[angle=0,width=12.50cm]{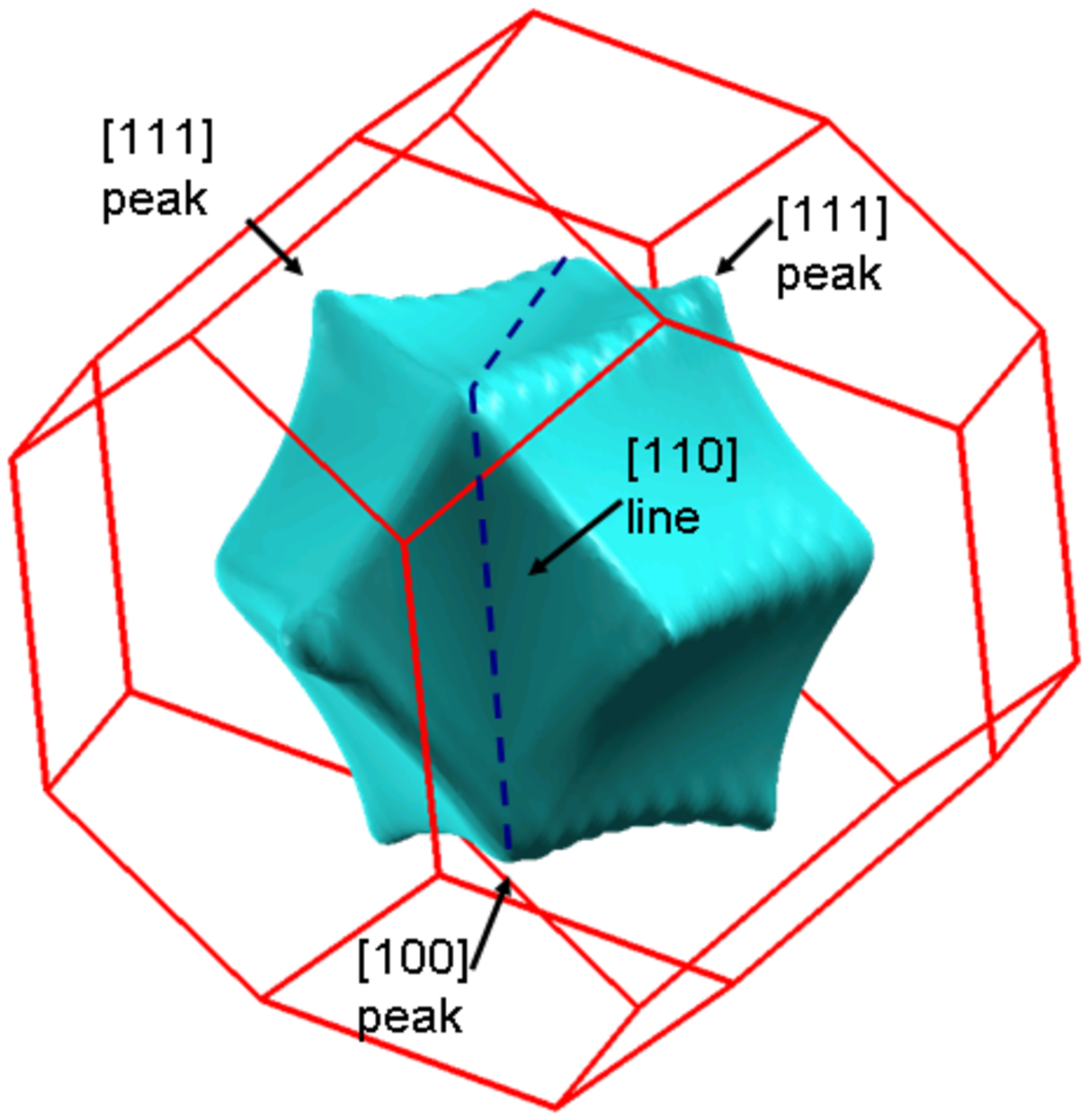}
\caption{}
\label{pd_fcc_sixth} 
\end{center} 
\end{figure}

\begin{figure}
\begin{center}
\centering
\includegraphics[angle=0,width=12.50cm]{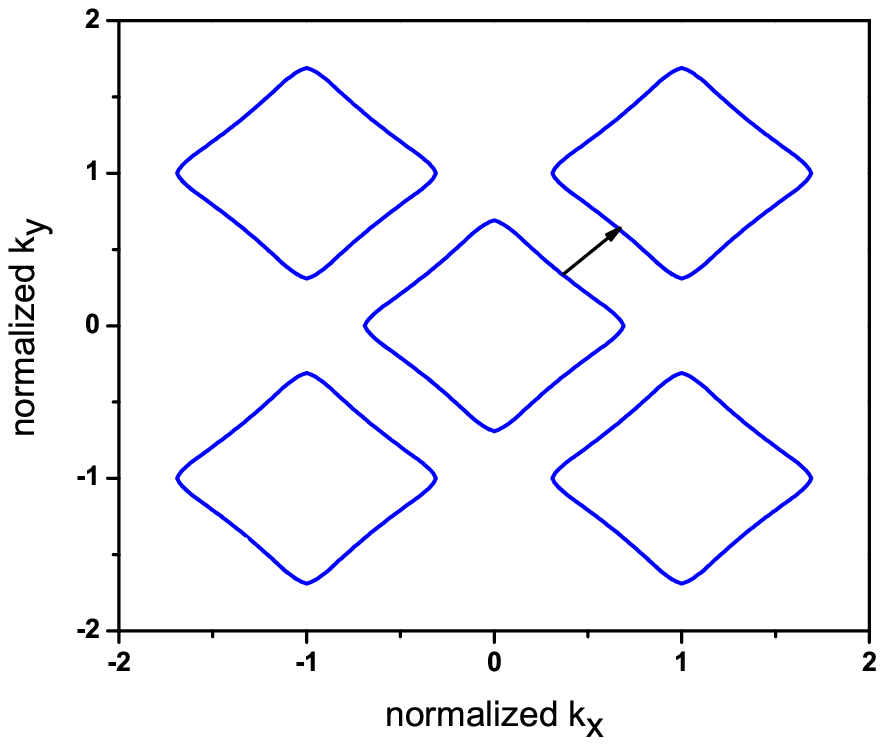}]
\caption{}
\label{kohn_anomaly} 
\end{center} 
\end{figure}

Based on analysis of the Fermi surface, I found that the $s$ character sixth band also provides contributions to the Kohn anomaly in the [110] direction.  As shown in Figure ~\ref{pd_fcc_sixth}, the sixth band Fermi surface has fairly wide flat valleys in the [110] directions that are essential for nesting vectors and Kohn anomalies.  The portions of the sixth band that intersect the $k_{z}=0$ plane are marked with blue lines for the center Brillouin zone and neighboring ones in Figure ~\ref{kohn_anomaly}.  In the [110] directions, these cross-sections have almost no curvature, making them ideal for Fermi nesting.  A representative nesting vector based on the Fermi surface between adjacent Brillouin zones is denoted by a black arrow.  The average nesting vector based on the Fermi surface in the [$k_{x},k_{y}$] plane is given by $\bf{q}=\frac{2\pi}{a}$[0.30, 0.30, 0] and is in good agreement with the Kohn anomaly calculated from density functional perturbation theory.  The observed Kohn anomaly is fairly broad due to the fact that parallel segments of the Fermi sheets off the [$k_{x},k_{y}$] plane also contribute to the anomaly at slightly different $q$ vectors.


The phonon frequencies along the [111] direction were calculated directly for both the transverse and longitudinal branches in an effort to resolve the phonon anomaly predicted by Freeman \textit{et al.}\cite{freeman1979}.  However, in this case, there was no clear evidence of a phonon anomaly in either acoustic branch.

\section{Conclusion}
In this work, I have considered the lattice dynamics of fcc palladium based on density functional perturbation theory.  Agreement between direct phonon calculations along symmetry lines with current experimental data for the full phonon spectrum of palladium is good.  In particular, I was able to reproduce the extended phonon anomaly in the [110] direction that had been observed experimentally.  In addition, I found no evidence for phonon anomalies in other branches of the phonon dispersion.  I also examined the phonon dispersion derived from interatomic force constants and looked at the influence of $q$-vector sampling.  The interatomic force constant approach with a 8x8x8 $q$-vector grid was able to provide good agreement for general features of the palladium phonon dispersion.

\begin{acknowledgments}
Calculations were performed on the Intel Cluster at the Cornell Nanoscale Facility which is part of the National Nanotechnology Infrastructure Network (NNIN) funded by the National Science Foundation.  Fermi surface images were generated using XCrySDens software package\cite{kokalj}.  I also wish to thank Eyvaz Isaev for providing the Fermi surface conversion utilities for Quantum Espresso.
\end{acknowledgments}


\begin{thebibliography}{27}
\expandafter\ifx\csname natexlab\endcsname\relax\def\natexlab#1{#1}\fi
\expandafter\ifx\csname bibnamefont\endcsname\relax
  \def\bibnamefont#1{#1}\fi
\expandafter\ifx\csname bibfnamefont\endcsname\relax
  \def\bibfnamefont#1{#1}\fi
\expandafter\ifx\csname citenamefont\endcsname\relax
  \def\citenamefont#1{#1}\fi
\expandafter\ifx\csname url\endcsname\relax
  \def\url#1{\texttt{#1}}\fi
\expandafter\ifx\csname urlprefix\endcsname\relax\def\urlprefix{URL }\fi
\providecommand{\bibinfo}[2]{#2}
\providecommand{\eprint}[2][]{\url{#2}}

\bibitem[{\citenamefont{Favier et~al.}(2001)\citenamefont{Favier, Walter, Zach,
  Benter, and Penner}}]{hydrogen_sensor_2001}
\bibinfo{author}{\bibfnamefont{F.}~\bibnamefont{Favier}},
  \bibinfo{author}{\bibfnamefont{E.~C.} \bibnamefont{Walter}},
  \bibinfo{author}{\bibfnamefont{M.~P.} \bibnamefont{Zach}},
  \bibinfo{author}{\bibfnamefont{T.}~\bibnamefont{Benter}}, \bibnamefont{and}
  \bibinfo{author}{\bibfnamefont{R.~M.} \bibnamefont{Penner}},
  \bibinfo{journal}{Science} \textbf{\bibinfo{volume}{293}},
  \bibinfo{pages}{2227} (\bibinfo{year}{2001}).

\bibitem[{\citenamefont{Javey et~al.}(2003)\citenamefont{Javey, Guo, Wang,
  Lundstrom, and Dai}}]{javey}
\bibinfo{author}{\bibfnamefont{A.}~\bibnamefont{Javey}},
  \bibinfo{author}{\bibfnamefont{J.}~\bibnamefont{Guo}},
  \bibinfo{author}{\bibfnamefont{Q.}~\bibnamefont{Wang}},
  \bibinfo{author}{\bibfnamefont{M.}~\bibnamefont{Lundstrom}},
  \bibnamefont{and} \bibinfo{author}{\bibfnamefont{H.}~\bibnamefont{Dai}},
  \bibinfo{journal}{Nature} \textbf{\bibinfo{volume}{424}},
  \bibinfo{pages}{654} (\bibinfo{year}{2003}).

\bibitem[{\citenamefont{Baroni et~al.}(1987)\citenamefont{Baroni, Giannozzi,
  and Testa}}]{baroni1987}
\bibinfo{author}{\bibfnamefont{S.}~\bibnamefont{Baroni}},
  \bibinfo{author}{\bibfnamefont{P.}~\bibnamefont{Giannozzi}},
  \bibnamefont{and} \bibinfo{author}{\bibfnamefont{A.}~\bibnamefont{Testa}},
  \bibinfo{journal}{Phys. Rev. Lett.} \textbf{\bibinfo{volume}{58}},
  \bibinfo{pages}{1861} (\bibinfo{year}{1987}).

\bibitem[{\citenamefont{Baroni et~al.}(2001)\citenamefont{Baroni, de~Gironcoli,
  and Corso}}]{dfpt_review}
\bibinfo{author}{\bibfnamefont{S.}~\bibnamefont{Baroni}},
  \bibinfo{author}{\bibfnamefont{S.}~\bibnamefont{de~Gironcoli}},
  \bibnamefont{and} \bibinfo{author}{\bibfnamefont{A.~D.} \bibnamefont{Corso}},
  \bibinfo{journal}{Rev. Mod. Phys.} \textbf{\bibinfo{volume}{73}},
  \bibinfo{pages}{515} (\bibinfo{year}{2001}).

\bibitem[{\citenamefont{Miiller and Brockhouse}(1968)}]{brockhouse1968}
\bibinfo{author}{\bibfnamefont{A.~P.} \bibnamefont{Miiller}} \bibnamefont{and}
  \bibinfo{author}{\bibfnamefont{B.~N.} \bibnamefont{Brockhouse}},
  \bibinfo{journal}{Phys. Rev. Lett.} \textbf{\bibinfo{volume}{20}},
  \bibinfo{pages}{798} (\bibinfo{year}{1968}).

\bibitem[{\citenamefont{Miiller and Brockhouse}(1971)}]{brockhouse1971}
\bibinfo{author}{\bibfnamefont{A.~P.} \bibnamefont{Miiller}} \bibnamefont{and}
  \bibinfo{author}{\bibfnamefont{B.~N.} \bibnamefont{Brockhouse}},
  \bibinfo{journal}{Can. J. Phys.} \textbf{\bibinfo{volume}{49}},
  \bibinfo{pages}{704} (\bibinfo{year}{1971}).

\bibitem[{\citenamefont{Miiller}(1975)}]{miiller1975}
\bibinfo{author}{\bibfnamefont{A.~P.} \bibnamefont{Miiller}},
  \bibinfo{journal}{Can. J. Phys.} \textbf{\bibinfo{volume}{53}},
  \bibinfo{pages}{2492} (\bibinfo{year}{1975}).

\bibitem[{\citenamefont{Kohn}(1959)}]{kohn}
\bibinfo{author}{\bibfnamefont{W.}~\bibnamefont{Kohn}}, \bibinfo{journal}{Phys.
  Rev. Lett.} \textbf{\bibinfo{volume}{2}}, \bibinfo{pages}{393}
  (\bibinfo{year}{1959}).

\bibitem[{\citenamefont{Freeman et~al.}(1979)\citenamefont{Freeman,
  Watson-Yang, and Rath}}]{freeman1979}
\bibinfo{author}{\bibfnamefont{A.~J.} \bibnamefont{Freeman}},
  \bibinfo{author}{\bibfnamefont{T.~J.} \bibnamefont{Watson-Yang}},
  \bibnamefont{and} \bibinfo{author}{\bibfnamefont{J.}~\bibnamefont{Rath}},
  \bibinfo{journal}{J. Magn. Magn. Mater.} \textbf{\bibinfo{volume}{12}},
  \bibinfo{pages}{140} (\bibinfo{year}{1979}).

\bibitem[{\citenamefont{Maliszewski et~al.}(1979)\citenamefont{Maliszewski,
  Bednarski, Czachor, and Sosnowski}}]{maliszewski1979}
\bibinfo{author}{\bibfnamefont{E.}~\bibnamefont{Maliszewski}},
  \bibinfo{author}{\bibfnamefont{S.}~\bibnamefont{Bednarski}},
  \bibinfo{author}{\bibfnamefont{A.}~\bibnamefont{Czachor}}, \bibnamefont{and}
  \bibinfo{author}{\bibfnamefont{J.}~\bibnamefont{Sosnowski}},
  \bibinfo{journal}{J. Phys. F. Metal Phys.} \textbf{\bibinfo{volume}{9}},
  \bibinfo{pages}{2335} (\bibinfo{year}{1979}).

\bibitem[{\citenamefont{Fielek}(1980)}]{fielek1980}
\bibinfo{author}{\bibfnamefont{B.~L.} \bibnamefont{Fielek}},
  \bibinfo{journal}{J. Phys. F: Met. Phys.} \textbf{\bibinfo{volume}{10}},
  \bibinfo{pages}{2381} (\bibinfo{year}{1980}).

\bibitem[{\citenamefont{Mohammed et~al.}(1984)\citenamefont{Mohammed, Shukla,
  Milstein, and Merz}}]{merz1984}
\bibinfo{author}{\bibfnamefont{K.}~\bibnamefont{Mohammed}},
  \bibinfo{author}{\bibfnamefont{M.~M.} \bibnamefont{Shukla}},
  \bibinfo{author}{\bibfnamefont{F.}~\bibnamefont{Milstein}}, \bibnamefont{and}
  \bibinfo{author}{\bibfnamefont{J.~L.} \bibnamefont{Merz}},
  \bibinfo{journal}{Phys. Rev. B} \textbf{\bibinfo{volume}{29}},
  \bibinfo{pages}{3117} (\bibinfo{year}{1984}).

\bibitem[{\citenamefont{Gupta}(1985)}]{Gupta1985}
\bibinfo{author}{\bibfnamefont{O.~P.} \bibnamefont{Gupta}},
  \bibinfo{journal}{J. Chem. Phys.} \textbf{\bibinfo{volume}{82}},
  \bibinfo{pages}{927} (\bibinfo{year}{1985}).

\bibitem[{\citenamefont{Antonov et~al.}(1991)\citenamefont{Antonov,
  Zhalko-Titarenko, Milman, Khotkevich, and Krainyukov}}]{antonov1991}
\bibinfo{author}{\bibfnamefont{V.~N.} \bibnamefont{Antonov}},
  \bibinfo{author}{\bibfnamefont{A.~V.} \bibnamefont{Zhalko-Titarenko}},
  \bibinfo{author}{\bibfnamefont{V.~Y.} \bibnamefont{Milman}},
  \bibinfo{author}{\bibfnamefont{A.~V.} \bibnamefont{Khotkevich}},
  \bibnamefont{and} \bibinfo{author}{\bibfnamefont{S.~N.}
  \bibnamefont{Krainyukov}}, \bibinfo{journal}{J. Phys. Condens. Matter}
  \textbf{\bibinfo{volume}{3}}, \bibinfo{pages}{6523} (\bibinfo{year}{1991}).

\bibitem[{\citenamefont{Savrasov and Savrasov}(1996)}]{savrasov1996}
\bibinfo{author}{\bibfnamefont{S.~Y.} \bibnamefont{Savrasov}} \bibnamefont{and}
  \bibinfo{author}{\bibfnamefont{D.~Y.} \bibnamefont{Savrasov}},
  \bibinfo{journal}{Phys. Rev. B} \textbf{\bibinfo{volume}{54}},
  \bibinfo{pages}{16487} (\bibinfo{year}{1996}).

\bibitem[{\citenamefont{Takezawa et~al.}(2005)\citenamefont{Takezawa, Nagara,
  and Suzuki}}]{takezawa}
\bibinfo{author}{\bibfnamefont{T.}~\bibnamefont{Takezawa}},
  \bibinfo{author}{\bibfnamefont{H.}~\bibnamefont{Nagara}}, \bibnamefont{and}
  \bibinfo{author}{\bibfnamefont{N.}~\bibnamefont{Suzuki}},
  \bibinfo{journal}{Phys. Rev. B} \textbf{\bibinfo{volume}{71}},
  \bibinfo{pages}{012515} (\bibinfo{year}{2005}).

\bibitem[{\citenamefont{Giannozzi et~al.}(1991)\citenamefont{Giannozzi,
  de~Gironcoli, Pavone, and Baroni}}]{giannozzi}
\bibinfo{author}{\bibfnamefont{P.}~\bibnamefont{Giannozzi}},
  \bibinfo{author}{\bibfnamefont{S.}~\bibnamefont{de~Gironcoli}},
  \bibinfo{author}{\bibfnamefont{P.}~\bibnamefont{Pavone}}, \bibnamefont{and}
  \bibinfo{author}{\bibfnamefont{S.}~\bibnamefont{Baroni}},
  \bibinfo{journal}{Phys. Rev. B} \textbf{\bibinfo{volume}{43}},
  \bibinfo{pages}{7231} (\bibinfo{year}{1991}).

\bibitem[{\citenamefont{Baroni and \textit{et al.}}()}]{pwscf}
\bibinfo{author}{\bibfnamefont{S.}~\bibnamefont{Baroni}} \bibnamefont{and}
  \bibinfo{author}{\bibnamefont{\textit{et al.}}},
  \urlprefix\url{http://www.pwscf.org}.

\bibitem[{\citenamefont{Perdew et~al.}(1996)\citenamefont{Perdew, Burke, and
  Ernzerhof}}]{pbe_gga}
\bibinfo{author}{\bibfnamefont{J.~P.} \bibnamefont{Perdew}},
  \bibinfo{author}{\bibfnamefont{K.}~\bibnamefont{Burke}}, \bibnamefont{and}
  \bibinfo{author}{\bibfnamefont{M.}~\bibnamefont{Ernzerhof}},
  \bibinfo{journal}{Phys. Rev. Lett.} \textbf{\bibinfo{volume}{77}},
  \bibinfo{pages}{3865} (\bibinfo{year}{1996}).

\bibitem[{\citenamefont{Perdew and Zunger}(1981)}]{pz_lda}
\bibinfo{author}{\bibfnamefont{J.~P.} \bibnamefont{Perdew}} \bibnamefont{and}
  \bibinfo{author}{\bibfnamefont{A.}~\bibnamefont{Zunger}},
  \bibinfo{journal}{Phys. Rev. B} \textbf{\bibinfo{volume}{23}},
  \bibinfo{pages}{5048} (\bibinfo{year}{1981}).

\bibitem[{\citenamefont{Chen et~al.}(1989)\citenamefont{Chen, Brener, and
  Callaway}}]{fcc_pd_mag_trans}
\bibinfo{author}{\bibfnamefont{H.}~\bibnamefont{Chen}},
  \bibinfo{author}{\bibfnamefont{N.~E.} \bibnamefont{Brener}},
  \bibnamefont{and} \bibinfo{author}{\bibfnamefont{J.}~\bibnamefont{Callaway}},
  \bibinfo{journal}{Phys. Rev. B} \textbf{\bibinfo{volume}{40}},
  \bibinfo{pages}{1443} (\bibinfo{year}{1989}).

\bibitem[{\citenamefont{Alexandre et~al.}(2006)\citenamefont{Alexandre,
  Mattesini, Soler, and Yndurain}}]{Alexandre2006}
\bibinfo{author}{\bibfnamefont{S.~S.} \bibnamefont{Alexandre}},
  \bibinfo{author}{\bibfnamefont{M.}~\bibnamefont{Mattesini}},
  \bibinfo{author}{\bibfnamefont{J.~M.} \bibnamefont{Soler}}, \bibnamefont{and}
  \bibinfo{author}{\bibfnamefont{F.}~\bibnamefont{Yndurain}},
  \bibinfo{journal}{Phys. Rev. Lett.} \textbf{\bibinfo{volume}{96}},
  \bibinfo{pages}{079701} (\bibinfo{year}{2006}).

\bibitem[{Sob()}]{Sob2005}
\bibinfo{note}{T. Kana, D. Legut, M. Sob, \textit{private communication}}.

\bibitem[{\citenamefont{Monkhorst and Pack}(1976)}]{mp_grid}
\bibinfo{author}{\bibfnamefont{H.~J.} \bibnamefont{Monkhorst}}
  \bibnamefont{and} \bibinfo{author}{\bibfnamefont{J.~D.} \bibnamefont{Pack}},
  \bibinfo{journal}{Phys. Rev. B} \textbf{\bibinfo{volume}{13}},
  \bibinfo{pages}{5188} (\bibinfo{year}{1976}).

\bibitem[{\citenamefont{Methfessel and Paxton}(1989)}]{methfessel_paxton1989}
\bibinfo{author}{\bibfnamefont{M.}~\bibnamefont{Methfessel}} \bibnamefont{and}
  \bibinfo{author}{\bibfnamefont{A.~T.} \bibnamefont{Paxton}},
  \bibinfo{journal}{Phys. Rev. B} \textbf{\bibinfo{volume}{40}},
  \bibinfo{pages}{3616} (\bibinfo{year}{1989}).

\bibitem[{\citenamefont{Piscanec et~al.}(2004)\citenamefont{Piscanec, Lazzeri,
  Mauri, Ferrari, and Robertson}}]{piscanec}
\bibinfo{author}{\bibfnamefont{S.}~\bibnamefont{Piscanec}},
  \bibinfo{author}{\bibfnamefont{M.}~\bibnamefont{Lazzeri}},
  \bibinfo{author}{\bibfnamefont{F.}~\bibnamefont{Mauri}},
  \bibinfo{author}{\bibfnamefont{A.~C.} \bibnamefont{Ferrari}},
  \bibnamefont{and}
  \bibinfo{author}{\bibfnamefont{J.}~\bibnamefont{Robertson}},
  \bibinfo{journal}{Phys. Rev. Lett.} \textbf{\bibinfo{volume}{93}},
  \bibinfo{pages}{185503} (\bibinfo{year}{2004}).

\bibitem[{\citenamefont{Kokalj}(2003)}]{kokalj}
\bibinfo{author}{\bibfnamefont{A.}~\bibnamefont{Kokalj}},
  \bibinfo{journal}{Comp. Mater. Sci.} \textbf{\bibinfo{volume}{28}},
  \bibinfo{pages}{155} (\bibinfo{year}{2003}).

\end{thebibliography}

\newpage
\noindent \textbf{List of Captions}

\begin{description}
\item{\textbf{Figure 1}}: (color online) Phonon dispersion for fcc palladium along different symmetry lines.  The phonon dispersion based on the 8x8x8 q-vector grid interatomic force constants is given by black lines.  The results from the direct phonon calculations are denoted by black triangles and shown in comparison with experimental results from Miiller and Brockhouse\cite{brockhouse1971} taken at 120 K (red circles).

\item{\textbf{Figure 2}}: (color online) Phonon dispersion for the TA$_1$ branch of fcc Pd along the [$\zeta\zeta 0$] line from $\Gamma$ to X.  The direct calculations (red triangles) and IFC-based results (blue line) are shown in comparison with the experimental results from Miiller\cite{miiller1975} taken at 8 K (black circles).  The location of the phonon anomaly is denoted by the arrow on the graph.  

\item{\textbf{Figure 3}}: (color online) Phonon dispersion for the TA$_1$ branch of fcc Pd is shown along the [$\zeta\zeta 0$] line from $\Gamma$ to X for different values of the smearing parameter.  The results from calculations using $\sigma=0.01$ Ry (open diamonds), $\sigma=0.05$ Ry (black squares), and $\sigma=0.10$ Ry (red circles) are shown in comparison with the experimental results from Miiller\cite{miiller1975} taken at 8 K (triangles).

\item{\textbf{Figure 4}}: (color online) The Fermi surface as viewed head on from the [100] direction is shown in (a).  The Fermi surface for the fourth band forms small cusps in the [100] faces and is shown in dark blue.  The Fermi surface for the fifth band is denoted by a purple interior and green exterior and forms petals center on the [100] faces.  The sixth band with high $s$ character is located at the center of the Brillouin zone and is shown in light blue.  In part (b), the Fermi surface for the fifth band is shown alone for greater clarity.

\item{\textbf{Figure 5}}: (color online) The intersection of the fifth band Fermi surface (orange line) through the $k_{z}=0$ plane is shown.  The Fermi surface intersections in relevant neighboring Brillouin zones are also shown (blue lines).  The nesting vectors, $\bf{q}=\frac{2\pi}{a}$[0.30, 0.30, 0] between adjacent sections of the Fermi surface is denoted by black arrows.

\item{\textbf{Figure 6}}: (color online) The Fermi surface for the sixth band is shown in light blue within the fcc Brillouin zone (red lines).  This band has a high $s$ character and peaks in the [111] and [100] directions.  A dashed line indicates the center of the wide flat valleys in the [110] directions.  These regions are crucial for the nesting vectors which generate the Kohn anomaly in the [110] direction.

\item{\textbf{Figure 7}}: (color online) A cross-section of the sixth band Fermi surface (blue lines) through the $k_{z}=0$ plane is shown.  The Fermi surface cross-sections from neighboring Brillouin zones are also shown.  The nesting vector, $\bf{q}=\frac{2\pi}{a}$[0.30, 0.30, 0] between adjacent sections of the Fermi surface is denoted by a black arrow.
\end{description}

\clearpage
\clearpage

\clearpage
\clearpage

\clearpage
\clearpage

\clearpage
\clearpage

\clearpage
\clearpage

\clearpage
\clearpage

\clearpage

\end{document}